\documentclass{jnmp}

\begin{document}

\renewcommand{\evenhead}{O Lechtenfeld and A Sorin}
\renewcommand{\oddhead}{$N=(1|1)$ Supersymmetric Toda Lattice Hierarchy}

\setcounter{page}{183}
\thispagestyle{empty}


\FirstPageHead{8}{2}{2001}
{\pageref{Lech-firstpage}--\pageref{Lech-lastpage}}{Letter}

\copyrightnote{2001}{O Lechtenfeld and A Sorin}

\Name{Hidden $\pbf{N=(2|2)}$ Supersymmetry of the\\ 
$\pbf{N=(1|1)}$ Supersymmetric 
Toda Lattice Hierarchy}\label{Lech-firstpage}

\Author{O. LECHTENFELD~$^{\dag}$ and A SORIN~$^{\ddag}$}

\Address{$^{\dag }$~Institut f\"ur Theoretische Physik,
Universit\"at Hannover\\
~~Appelstra\ss{}e 2, D-30167 Hannover, Germany\\
~~E-mail: lechtenf@itp.uni-hannover.de\\[2mm]
$^{\ddag}$~Bogoliubov Laboratory of Theoretical Physics, JINR\\
~~141980 Dubna, Moscow Region, Russia\\
~~E-mail: sorin@thsun1.jinr.ru}

\Date{Received September 18, 2000; Accepted  December 10, 2000}

\begin{abstract}
\noindent
An $N=(2|2)$ superfield formulation of the $N=(1|1)$
supersymmetric Toda lattice hierarchy is proposed, and
its five real forms are presented.
\end{abstract}

\section{Introduction}

Recently the $N=(1|1)$ supersymmetric generalization 
\cite{ls} of the Darboux transformation~\cite{darboux} was proposed, 
and an infinite class 
of bosonic and fermionic solutions of its symmetry equation was 
constructed in \cite{ls,ols}. These solutions generate bosonic and 
fermionic flows of the $N=(1|1)$ supersymmetric Toda lattice hierarchy 
in the same way as their bosonic counterparts --- the solutions of the 
symmetry equation of the Darboux transformation \cite{dly} --- produce the 
flows of the bosonic Toda lattice hierarchy. Actually, the $N=(1|1)$ 
Toda lattice hierarchy is $N=(2|2)$ supersymmetric (see Section~3),
and henceforth we shall call it the $N=(2|2)$ Toda lattice hierarchy.
Naturally, the quest for its $N=(2|2)$ superfield formulation arises.

The present letter addresses this problem. In Section~2 we present a short
summary of the main facts concerning the $N=(2|2)$ Toda lattice
hierarchy and its bosonic and fermionic flows which are used in what
follows. In Section 3 we formulate a {\it conjecture\/} concerning the
$N=(2|2)$ superfield formulation of the $N=(2|2)$ Toda lattice hierarchy.
Our conjecture is partly proven in Section~4 and gains further support
in Section 5 by a set of arguments, including explicit calculations of the
first three flows. In Section~6 we also present five complex conjugations
in $N=(2|2)$ superspace which are admitted by the flows.

\section{$\pbf{N=(2|2)}$ Toda lattice hierarchy in $\pbf{N=(1|1)}$ superspace}

In this section we briefly review the approach of refs.~\cite{ls,ols}
(for more detail, see \cite{ls,ols} and references therein)
for constructing an infinite class of bosonic and fermionic flows
of the $N=(2|2)$ Toda lattice hierarchy in $N=(1|1)$ superspace.

The starting point is the $N=(1|1)$ supersymmetric generalization
of the Darboux transformation \cite{ls},
\begin{equation}
u_{j+1}= \frac{1}{v_j},\qquad
v_{j+1} = v_j (D_-D_+\ln v_j-u_jv_j),
\label{toda1}
\end{equation}
where $u_j\equiv u_j(x^+,\theta^+;x^-,\theta^-)$ and
$v_j\equiv v_j(x^+,\theta^+;x^-,\theta^-)$ are bosonic $N=(1|1)$
superfields defined on the lattice, $j \in {\mathbb Z}$,
and $D_{\pm}$ are the $N=1$ supersymmetric fermionic covariant derivatives
\begin{equation}
 D_{\pm} = \frac{\partial}{\partial \theta^{\pm}}+
\theta^{\pm}\frac{{\partial}}{{\partial x^{\pm}}},
\qquad D^2_{\pm} = \frac{{\partial}}{{\partial x^{\pm}}}
 \equiv {\partial}_{\pm},\qquad \{D_+,D_-\} = 0.
\label{sup}
\end{equation}
The composite superfield
\begin{equation}
b_j \equiv u_j v_j
\label{b}
\end{equation}
satisfies the $N=(1|1)$ supersymmetric Toda lattice equation
\begin{equation}
D_-D_+ \ln b_j = b_{j+1}-b_{j-1}.
\label{toda}
\end{equation}
For this reason, the hierarchy of equations invariant under the
Darboux transformation~(\ref{toda1}) is called the $N=(1|1)$
supersymmetric Toda lattice hierarchy.

\renewcommand{\footnoterule}{\vspace*{3pt}%
\noindent
\rule{.4\columnwidth}{0.4pt}\vspace*{6pt}}

One of the possible ways of constructing invariant equations is to solve
the corresponding symmetry equation. In the case under consideration it reads
\begin{equation}
U_{j+1} = -\frac{1}{v_{j}^2}V_j, \qquad
V_{j+1} = \frac{v_{j+1}}{v_j}V_j+v_j\left(D_-D_+
\left(\frac{1}{v_j}V_j\right)-v_jU_j-u_jV_j\right),
\label{sym}
\end{equation}
where $V_j$ and $U_j$ are bosonic functionals of the superfields $v_j$ and
$u_j$. Any particular solution $V^{p}_{j}$, $U^{p}_{j}$ generates an
evolution system of equations involving only the superfields $v_j$ and
$u_j$ defined at the same lattice point $j$,
with respect to a bosonic evolution time $t_p$,
\begin{equation}
\frac{{\partial}}{\partial t_p} v_j = V^{p}_j,\qquad
\frac{{\partial}}{\partial t_p} u_j = U^{p}_j.
\label{evol}
\end{equation}
By construction\footnote{Let us recall that eq.~(\ref{sym}) is just a result of
differentiating eq.~(\ref{toda1}) with respect to the evolution time $t_p$.},
this system is invariant under the discrete transformation (\ref{toda1})
and, therefore, belongs to the hierarchy as defined above. In other words,
different solutions of the evolution system~(\ref{evol})
(which, actually, are given by pairs of superfields $\{v_j,
u_{j}\}$ with different values for $j$) are related by
the discrete Darboux transformation (\ref{toda1}). Altogether,
invariant evolution systems form a {\it differential\/} hierarchy, i.e.
a hierarchy of equations involving only superfields at a single lattice
point\footnote{In the case of the one- (two-) dimensional bosonic Toda lattice
the differential hierarchy coincides with the nonlinear Schr\"odinger
(Davey--Stewartson) hierarchy \cite{bx,lsy,dly}.}. In contrast, the discrete
lattice shift (the Darboux transformation), when added
to the differential hierarchy, generates the {\it discrete\/} $N=(1|1)$
supersymmetric Toda lattice hierarchy. Thus, the
discrete hierarchy appears as a collection of an infinite number of
isomorphic differential hierarchies~\cite{bx}.

The symmetry equation (\ref{sym}) represents a complicated nonlinear functional
equation, and its general solution is not known.
For a more complete understanding of the hierarchy structure
and its solutions it seems advantageous to know
as many solutions of eq.~(\ref{sym}) as possible.
Refs.~\cite{ls,ols} addressed this problem and derived a wide class of
bosonic as well as fermionic solutions.

First, the functionals $V_j$ and $U_j$ are consistently represented
in terms of a single bosonic functional $\alpha_{0,j}[u_j,v_j]$,
\begin{equation}
V_j = -v_j \alpha_{0,j}, \qquad U_j = u_j{\alpha}_{0,j-1},
\label{evo1}
\end{equation}
in terms of which the symmetry equation (\ref{sym}) becomes
\begin{equation}
D_-D_+\alpha_{0,j} = b_{j+1}\,({\alpha}_{0,j+1}-\alpha_{0,j}) +
b_j(\alpha_{0,j}-{\alpha}_{0,j-1}),
\label{sym1}
\end{equation}
where the superfield $b_j$ is defined by eq.~(\ref{b})
and constrained by eq.~(\ref{toda}).

Second, the following recursive chain of substitutions is introduced:
\begin{equation}
\alpha^{\pm}_{p,j} = \pm D^{-1}_{\mp}\left(b_{j+p+1}\alpha^{\pm}_{p+1,j}
+ (-1)^{p} b_j {\alpha}^{\pm}_{p+1,j-1}{}\right), \qquad p=0,1,2,\ldots,
\label{rec}
\end{equation}
where $\alpha^{\pm}_{2p,j}$ ($\alpha^{\pm}_{2p+1,j}$) are new bosonic
(fermionic) functionals of length dimension
\begin{equation}
[\alpha^\pm_{p,j}] = [\alpha^{\pm}_{0,j}]+\frac{p}{2},
\label{dimal}
\end{equation}
and the superscripts $+$ and $-$ mark two different series of
solutions to the symmetry equation (\ref{sym1}).
Equations~(\ref{rec}) can be used to express $\alpha^{\pm}_{0,j}$ in terms of
$\alpha^\pm_{p,j}$ for any chosen~$p$.
The following equation for $\alpha^{\pm}_{p,j}$,
\begin{equation}\arraycolsep=0em
\begin{array}{l}
\displaystyle {\pm}(-1)^{p}D_{\pm} \alpha^{\pm}_{p,j}+{\alpha}^{\pm}_{p,j}
D^{-1}_{\mp}\left(b_{j+p+1}+b_{j+p}- b_{j+1}-b_j\right)
\vspace{3mm}\\
\qquad  = D^{-1}_{\mp}\left(b_{j+p+1}{\alpha}^{\pm}_{p,j+1}+
(-1)^{p}(b_{j+p}-b_{j+1}) \alpha^{\pm}_{p,j}-
b_j\, {\alpha}^{\pm}_{p,j-1}\right),
\end{array}\label{induction}
\end{equation}
can easily be proved by induction.

We now describe the solutions of the equations
arising in this iterative process. It turns out that, at any given $p$,
the equation~(\ref{induction}) possesses a very simple solution for
$\alpha^{\pm}_{p,j}$, namely
\begin{equation}
\alpha^{\pm}_{p,j} = (-1)^{pj}{\epsilon}_p \qquad \Rightarrow \qquad
[\alpha^{\pm}_{p,j}] = 0,
\label{recsolb}
\end{equation}
where ${\epsilon}_p$ is a dimensionless fermionic (bosonic) constant
for odd (even) values of $p$. Therefore,
the recursive procedure may be entered at any chosen $p$ with the simple
initial value~(\ref{recsolb}),
which then generates a very non-trivial solution
$\alpha^{(p)\pm}_{0,j}$
for the functional ${\alpha}{}^{\pm}_{0,j}$ via~(\ref{rec}).
The latter, in turn, yields the flows via eqs.~(\ref{evol}) and~(\ref{evo1}).

Let us demonstrate in more detail how bosonic and fermionic flows
originate from this background.

For the bosonic functionals $\alpha^{\pm}_{2p,j}$
the recursive procedure may be started at any even step.
The corresponding $\alpha^{(2p)\pm}_{0,j}$, being expressed in terms
of $\alpha^{\pm}_{2p,j}$ (\ref{recsolb}) via relations (\ref{rec}), has the
following symbolic form~\cite{ly},
\begin{equation}
\alpha^{(2p)\pm}_{0,j} = \pm \left[\prod^{2p}_{k=1}\left(1-(-1)^{k}
e^{-\left(k{\partial}_k+\sum\limits^{2p}_{n=k+1}{\partial}_n\right)}
\right)\right]
\left(\prod^{2p}_{m=1} D^{-1}_{\mp}b_{j+m}\right),
\label{recsolution}
\end{equation}
and generates the
$p$-th bosonic flow of the hierarchy,
\begin{equation}
\frac{\partial}{\partial t^{\pm}_p} v_j = -v_j
{\alpha^{(2p)\pm}_{0,j}},\qquad
\frac{\partial}{\partial t^{\pm}_p}u_j = u_j
{\alpha}^{(2p)\pm}_{0,j-1}
\qquad\Rightarrow\qquad [t^{\pm}_p] = -[\alpha^{(2p)\pm}_{0,j}] =p, 
\label{evolb}
\end{equation}
where we have used eqs.~(\ref{evol}), (\ref{evo1}), 
(\ref{dimal}), and (\ref{recsolb}),
and the superscripts $+$ and $-$ correspond to the two different series
(\ref{rec}) of solutions to the symmetry equation (\ref{sym1}).
The operator $e^{-l{\partial}_k}$ ($l \in {\mathbb Z}$) is the discrete lattice
shift which acts in eq.~(\ref{recsolution}) according the rule
\begin{equation}
e^{-l{\partial}_k}
\displaystyle \left(\prod^{2p}_{m=1} D^{-1}_{\mp}b_{j+m}\right) 
:= \left(\prod^{k-1}_{m=1} D^{-1}_{\mp}b_{j+m}\right) D^{-1}_{\mp}b_{j+k-l}
\left(\prod^{2p}_{m=k+1} D^{-1}_{\mp}b_{j+m}\right),
\label{rule1}
\end{equation}
i.e. ${\partial}_k$ is meant to act only on $b_{j+k}$ in the product.
By definition, the lattice shift operator
$e^{-l{\partial}_k}$ commutes with the fermionic covariant derivatives
$D_{\pm}$,
\begin{equation}
[e^{-l{\partial}_k}, D_{\pm}]=0.
\label{comrulshift}
\end{equation}
Although the solution $\alpha^{(2p)\pm}_{0,j}$ depends on all
superfields $v_{j+k}$ and $u_{j+k}$ with $0 \leq k \leq 2p$,
by using eq.~(\ref{toda1}) it can be expressed completely in terms of
the superfields $u_j$ and $v_j$ defined at the single lattice point $j$.
In this way the differential hierarchy of bosonic flows~(\ref{evolb})
 is generated (see the discussion after eq.~(\ref{evol})).
For illustration, we present the first two~\cite{ls}:
\begin{equation}
\frac{{\partial}}{{\partial t^{+}_1}} v = v, \qquad
\frac{{\partial}}{{\partial t^{+}_1}} u = u,
\label{bf1}
\end{equation}
\begin{equation}\arraycolsep=0em
\begin{array}{l}
\displaystyle \frac{{\partial}}{{\partial t^{+}_2}} v =
+{\partial}^{2}_+ v-2(D_+v)D^{-1}_-{\partial}_+(uv) +
2vD^{-1}_-\left[{\partial}_+(vD_+u)+2uvD^{-1}_-{\partial}_+(uv)\right],
\vspace{3mm}\\
\displaystyle
\frac{{\partial}}{{\partial t^{+}_2}} u =
-{\partial}^{2}_+ u -2(D_+u)D^{-1}_-{\partial}_+(uv)
+2uD^{-1}_-\left[{\partial}_+ (uD_+v)- 2uvD^{-1}_-{\partial}_+(uv)\right],
\end{array}
\label{bf2}
\end{equation}
where $u\equiv u_j(x^+,\theta^+;x^-,\theta^-)$ and
$v\equiv v_j(x^+,\theta^+;x^-,\theta^-)$.

For the fermionic functionals $\alpha^{\pm}_{2p-1,j}$
the recursive procedure may be started at any odd step.
It remains to show how fermionic flows are being activated.
This goal in mind, let us represent the bosonic time derivative entering
eq.~(\ref{evol}) in the following form:
\begin{equation}
\frac{{\partial}}{\partial t_p} =
{\epsilon}_{2p-1} D_p,
\label{evolder}
\end{equation}
defining a fermionic time derivative $D_p$. Then, eq.~(\ref{evol}) becomes
\begin{equation}\arraycolsep=0em
\begin{array}{l}
{\epsilon}_{2p-1} D^{\pm}_p v_j
= -v_j \alpha^{(2p-1)\pm}_{0,j},\quad
{\epsilon}_{2p-1} D^{\pm}_p u_j
= u_j \alpha^{(2p-1)\pm}_{0,j-1}  
\vspace{2mm}\\
\qquad \qquad \Rightarrow \quad
[D^{\pm}_p] = [\alpha^{(2p-1)\pm}_{0,j}]
= -p{+}{\textstyle{\frac{1}{2}}}, 
\end{array}
\label{evolf}
\end{equation}
where $\alpha^{(2p-1)\pm}_{0,j}$ should be expressed in terms
of $\alpha^{\pm}_{2p-1,j}$ (\ref{recsolb}) via relations~(\ref{rec}),
and eqs.~(\ref{evo1}), (\ref{dimal}) and (\ref{recsolb}) have been exploited
to arrive at eqs.~(\ref{evolf}).
The superscripts on $D^{\pm}_p$ in eqs.~(\ref{evolf}) again correspond to the 
two different series~(\ref{rec}) of solutions to the symmetry 
equation~(\ref{sym1}).
The fermionic constant
${\epsilon}_{2p-1}$ enters linearly 
both sides of eqs.~(\ref{evolf}), hence
the fermionic flows $D^{\pm}_p$ actually do not depend on
${\epsilon}_{2p-1}$.  In this context we remark that ${\epsilon}_{2p-1}$ is
an artificial parameter which need not be introduced at all. However, without
${\epsilon}_{2p-1}$ it is necessary to consider the quantities $t_p$,
$V^{p}_j$, $U^{p}_j$, $\alpha^{\pm}_{2p,j}$ ($\alpha^{\pm}_{2p-1,j}$)
entering eqs.~(\ref{evol}), (\ref{rec}) as fermionic (bosonic)
ones from the beginning. Of course, at the end of the analysis
one arrives at the same result~(\ref{evolf}).
For illustration, we present the first two fermionic flows from the set
(\ref{evolf}) \cite{ols}:
\begin{equation}
(-)^{j}D^{+}_1 v = -D_+v+2vD^{-1}_-(uv), \qquad
(-)^{j}D^{+}_1 u  = -D_+u-2uD^{-1}_-(uv),
\label{ff1}
\end{equation}
\begin{equation}\arraycolsep=0em
\begin{array}{l}
(-)^{j}D^{+}_2 v= -D_+{\partial}_+v+2({\partial}_+v)D^{-1}_-(uv)
\vspace{2mm}\\
\phantom{(-)^{j}D^{+}_2 u={}} + (D_+v)D^{-1}_-D_+(uv) + 
vD^{-1}_-[u{\partial}_+v + (D_+v)D_+u],
\vspace{2mm}\\
(-)^{j}D^{+}_2 u
= +D_+{\partial}_+u+2({\partial}_+u)D^{-1}_-(uv)
\vspace{2mm}\\
\phantom{(-)^{j}D^{+}_2 u={}} + (D_+u)D^{-1}_-D_+(uv) + uD^{-1}_-[v{\partial}_+u + (D_+u)D_+v].
\end{array}
\label{ff2}
\end{equation}

Let us note that the two differential hierarchies arising for the
two different values of $(-1)^{j}$ ($+1$ or $-1$) are actually
isomorphic. Indeed, one can easily see that they are related by the
standard automorphism
which changes the sign of all Grassmann numbers.
Thus, in distinction 
the bosonic Toda lattice, where the Darboux
transformation does not change the direction of evolution times in the
differential hierarchy~(\ref{evol}), 
its supersymmetric counterpart~(\ref{toda1})
reverses the sign of fermionic times in the differential hierarchy. This
supersymmetric peculiarity has no effect on the property
that the supersymmetric {\it discrete\/} hierarchy is a
collection of isomorphic {\it differential\/} hierarchies
like in the bosonic 
case\footnote{
For the one-dimensional bosonic Toda lattice hierarchy the isomorphism
which relates the differential hierarchies is trivial because
they are identical copies of the single nonlinear Schr\"odinger hierarchy
\cite{bx}.}.

The flows $D^{-}_k$ and $\frac{{\partial}}{\partial t^{-}_k}$ can easily be
derived by applying the invariance transformations
\begin{equation}
{\partial}_{\pm}\ \longrightarrow\ {\partial}_{\mp}, \qquad
D_{\pm}\ \longrightarrow\ \pm D_{\mp}
\label{supaut}
\end{equation}
of the $N=(1|1)$ supersymmetry algebra (\ref{sup}) and eqs.~(\ref{toda1}),
(\ref{toda}) and (\ref{sym1}) to the flows $D^{+}_k$ 
(\ref{ff1})--(\ref{ff2}) and
$\frac{{\partial}}{\partial t^{+}_k}$ (\ref{bf1})--(\ref{bf2}), respectively,
but we do not write them down here.

Using the explicit expressions for the constructed bosonic and fermionic
flows, one can calculate their algebra
\begin{equation}\arraycolsep=0em
\begin{array}{l}
\displaystyle 
\left\{D^{\pm}_k\,,\,D^{\pm}_l\right\} =
-2\frac{{\partial}}{{\partial t^{\pm}_{k+l-1}}},
\vspace{2mm}\\
\displaystyle \left\{D^{+}_k,D^{-}_l\right\} =
\left[\frac{{\partial}}{\partial t^{\pm}_k},
\frac{{\partial}}{\partial t^{\pm}_l}\right] =
\left[\frac{{\partial}}{\partial t^{+}_k},
\frac{{\partial}}{\partial t^{-}_l}\right] =
\left[\frac{{\partial}}{\partial t^{\pm}_k},
D^{\pm}_l\right] =
\left[\frac{{\partial}}{\partial t^{\pm}_k},
D^{\mp}_l\right] = 0,
\end{array}
\label{alg}
\end{equation}
which may be realized in the superspace  $\{t^{+}_k, \theta^{+}_k;
t^{-}_k,\theta^{-}_k\}$ via
\begin{equation}
D^{\pm}_k =
\frac{\partial}{\partial \theta^{\pm}_k}-
\sum^{\infty}_{l=1}\theta^{\pm}_l
\frac{\partial}{{\partial t^{\pm}_{k+l-1}}},
\label{covder}
\end{equation}
where $\theta^{+}_k$ and $\theta^{-}_k$ are abelian fermionic evolution
times.

\section{$\pbf{N=(2|2)}$ Toda lattice hierarchy in $\pbf{N=(2|2)}$ superspace}

In $N=(1|1)$ superspace, the additional supersymmetry of the
$N=(2|2)$ Toda lattice hierarchy is not manifest.
Yet, besides the two fermionic flows
$D^{\pm}_1$ in (\ref{ff1}) and (\ref{supaut}), there exist two more fermionic
flows $Q^{\pm}_1$. These are generated by the two obvious solutions of the
symmetry equation (\ref{sym}) which originate from the standard supersymmetric
transformations of the superfields,
\begin{equation}
Q^{\pm}_1 v = Q_{\pm}v, \qquad
Q^{\pm}_1 u  = Q_{\pm}u,
\label{supflows}
\end{equation}
where $Q_{\pm}$ are $N=(1|1)$ supersymmetric generators,
\begin{equation}\arraycolsep=0em
\begin{array}{l}
\displaystyle  Q_{\pm} = \frac{\partial}{\partial \theta^{\pm}}-
\theta^{\pm}\frac{{\partial}}{{\partial x^{\pm}}}, 
\qquad Q^2_{\pm} = -{\partial}_{\pm},\qquad \{Q_+,Q_-\} = 0,
\vspace{3mm}\\
\displaystyle  \{Q_+,D_{\pm}\} = 0,\qquad \{Q_-,D_{\pm}\} = 0.
\end{array}
\label{supgenerators}
\end{equation}
Altogether, the flows
$\left\{\frac{{\partial}}{\partial t^{\pm}_1},Q^{\pm}_1,D^{\pm}_1\right\}$
form the superalgebra of complex $N=(2|2)$ supersymmetry.
It will turn out that one of the real forms of the hierarchy realizes the
{\it real\/} $N=(2|2)$ supersymmetry algebra on its flows
(see the discussion after eq.~(\ref{conj2new})).

The existence of the hidden $N=(2|2)$ supersymmetry naturally raises the
problem of finding a very particular basis (if any), where it is realized
locally and linearly.
Its solution would correspond to constructing
an $N=(2|2)$ superfield formulation of the hierarchy.
With this aim in mind, it is instructive to rewrite the equations
(\ref{toda1}) and (\ref{evolb}) to the new superfield basis
$\{J_j, {\overline J}_j\}$,
\begin{equation}
{\overline J}_j :=  -u_jv_j \equiv -b_j, \qquad
J_j : = u_jv_j + D_-D_+\ln u_j,
\label{basis}
\end{equation}
which possesses the above-mentioned properties:
\begin{equation}\arraycolsep=0em
\begin{array}{l}
Q^{\pm}_1 J_j= Q_{\pm}J_j, \qquad
(-1)^j D^{\pm}_1 J_j = +D_{\pm}J_j,
\vspace{2mm}\\
Q^{\pm}_1  {\overline J}_j = Q_{\pm}{\overline J}_j, \qquad
(-1)^j D^{\pm}_1  {\overline J}_j  = -D_{\pm}{\overline J}_j,
\end{array}
\label{greatbasis}
\end{equation}
where eqs. (\ref{ff1}), (\ref{supaut}) and (\ref{basis}) have been used.
The new superfields $J_j\equiv J_j(x^+,\theta^+;x^-,\theta^-)$ and
${\overline J}_j\equiv {\overline J}_j(x^+,\theta^+;x^-,\theta^-)$ are
unconstrained bosonic $N=(1|1)$ lattice superfields.  They are related by
\begin{equation}
J_{j+1}= {\overline J}_j,\qquad
{\overline J}_{j+1} = J_j -D_-D_+\ln {\overline J}_j
\label{toda11}
\end{equation}
and satisfy
\begin{equation}
\frac{\partial}{\partial t^{\pm}_p}{\overline J}_j = {\overline J}_j
\left({\alpha^{(2p)\pm}_{0,j-1}}- {\alpha^{(2p)\pm}_{0,j}}\right),\qquad
\frac{\partial}{\partial t^{\pm}_p}J_j = J_j
\left({\alpha}{}^{(2p)\pm}_{0,j-2}-{\alpha^{(2p)\pm}_{0,j-1}}\right),
\label{evolb1}
\end{equation}
with $\alpha^{(2p)\pm}_{0,j}$ now being understood as functionals of
$J_j$ and ${\overline J}_j$.

At this point we formulate our {\it conjecture\/}.
We claim that the sought-for $N=(2|2)$ superspace formulation is achieved
simply by elevating the $N=(1|1)$ lattice superfields
$J_j$ and ${\overline J}_j$
to chiral resp. antichiral bosonic $N=(2|2)$ lattice superfields
${\cal J}_j(x^+,\theta^+,\eta^+;$ $x^-, \theta^-,\eta^-)$ and
${\overline {\cal J}}_j(x^+,\theta^+,\eta^+;x^-,\theta^-,\eta^-)$.
More concretely, the resulting equations
\begin{equation}
{\cal J}_{2(j+1)}=
{\cal J}_{2j} -{\cal D}_-{\cal D}_+\ln {\overline {\cal J}}_{2j},\qquad
{\overline {\cal J}}_{2(j+1)}={\overline {\cal J}}_{2j} -
{\overline {\cal D}}^-{\overline {\cal D}}^+
\ln {\cal J}_{2(j+1)}
\label{toda11r}
\end{equation}
and
\begin{equation}
\frac{\partial}{\partial t^{\pm}_p}{\overline {\cal J}}_j =
{\overline {\cal J}}_j
\left({\alpha^{(2p)\pm}_{0,j-1}}- {\alpha^{(2p)\pm}_{0,j}}\right),\qquad
\frac{\partial}{\partial t^{\pm}_p}{\cal J}_j = {\cal J}_j
\left({\alpha}{}^{(2p)\pm}_{0,j-2}-{\alpha^{(2p)\pm}_{0,j-1}}\right)
\label{evolb1r}
\end{equation}
are conjectured to be consistent with the chirality constraints
\begin{equation}
{\overline {\cal D}}^{\pm} {\overline {\cal J}}_{2j} =
{\cal D}_{\pm} {\overline {\cal J}}_{2j+1} = 0
\qquad{\rm and}\qquad
{\cal D}_{\pm}{\cal J}_{2j} =
{\overline {\cal D}}^{\pm}{\cal J}_{2j+1} =0.
\label{N=4constr}
\end{equation}
We would like to emphasize that the last statement is no trivial matter
because such a procedure in general leads to inconsistent equations
except for very special cases one of which is under consideration.
In the above, ${\cal D}_{\pm}$ and ${\overline {\cal D}}^{\pm}$ are
$N=(2|2)$ supersymmetric fermionic covariant derivatives,
\begin{equation}\arraycolsep=0em
\begin{array}{l}
\displaystyle {\cal D}_{\alpha} := \frac{1}{2}\left(
\frac{\partial}{\partial {\theta}^{{\alpha}}} +
i\frac{\partial}{\partial {\eta}^{{\alpha}}}+
\left({\theta}^{{\alpha}} +i{\eta}^{{\alpha}}\right) 
{\partial}_{{\alpha}}\right),
\vspace{3mm}\\
\displaystyle {\overline {\cal D}}^{{\alpha}} := \frac{1}{2}
\left(\frac{\partial}{\partial {\theta}^{{\alpha}}} -
i\frac{\partial}{\partial {\eta}^{{\alpha}}}+\left({\theta}^{{\alpha}} -
i{\eta}^{{\alpha}}\right) {\partial}_{{\alpha}}\right),
\qquad {\alpha},{\beta}=\pm, 
\vspace{3mm}\\
\displaystyle D_{\pm} = {\cal D}_{\pm}+ {\overline {\cal D}}^{\pm}, \qquad
\left\{{\cal D}_{{\alpha}}, {\overline {\cal D}}^{{\beta}}\right\}=
{{\delta}_{{\alpha}}}^{{\beta}} {\partial}_{{\beta}}, \qquad
\left\{{\cal D}_{{\alpha}}\,,\,{\cal D}_{{\beta}}\right\}=
\left\{{\overline {\cal D}}^{{\alpha}},
{\overline {\cal D}}^{{\beta}}\right\}=0,
\end{array}
\label{algnn4}
\end{equation}
and $\eta^{\pm}$ are two additional Grassmanian coordinates.
Since the right hand sides of eqs.~(\ref{evolb1r})
are solutions of the symmetry equation corresponding to
 eqs.~(\ref{toda11r}), we must require that the functionals
${\alpha^{\pm}_{0,j}} - {\alpha^{\pm}_{0,j-1}}$ entering eqs.~(\ref{evolb1r})
possess the following chirality properties:
\begin{equation}
{\overline {\cal D}}^{\mp}
\left({\alpha^{(2p)\pm}_{0,2j}} - {\alpha^{(2p)\pm}_{0,2j-1}}\right)=0, \qquad
{\cal D}_{\mp} \left({\alpha^{(2p)\pm}_{0,2j+1}} -
{\alpha^{(2p)\pm}_{0,2j}}\right)=0,
\label{constralpha}
\end{equation}
\begin{equation}
{\overline {\cal D}}^{\pm}
\left({\alpha^{(2p)\pm}_{0,2j}} - {\alpha^{(2p)\pm}_{0,2j-1}}\right)=0, \qquad
{\cal D}_{\pm} \left({\alpha^{(2p)\pm}_{0,2j+1}} -
{\alpha^{(2p)\pm}_{0,2j}}\right)=0.
\label{constralpha0}
\end{equation}
These four equations are necessary and sufficient conditions
for the consistency of~(\ref{toda11r}) and (\ref{evolb1r})
with the constraints~(\ref{N=4constr}).
Hence, we should set out to prove them.

\section{Proof of half the conjecture}

In the following,
we present a {\it proof\/} that the constraints~(\ref{constralpha})
are in fact satisfied. What concerns
the remaining constraints~(\ref{constralpha0}), we shall give evidence
in their favour in the next section, by confirming them
(and (\ref{constralpha})) explicitly for the first three flows from
the set~(\ref{evolb1r}).

First, the equations~(\ref{toda11r}) are obviously consistent with the
chirality constraints~(\ref{N=4constr}) and represent a manifestly $N=(2|2)$
supersymmetric form of the $N=(2|2)$ supersymmetric Toda lattice equations
(see, e.g. refs.~\cite{eh,lds} and references therein). From the chirality
constraints~(\ref{N=4constr}) one can derive the following intertwining
relations of the fermionic covariant derivatives ${\cal D}_{\pm}$ and
${\overline {\cal D}}^{\pm}$ with the lattice shift
operator $e^{\partial}$:
\begin{equation}
e^{\partial}{\overline {\cal D}}^{\mp}=
{\cal D}_{\mp}~e^{\partial}, \qquad
e^{\partial}~{\cal D}_{\mp}=
{\overline {\cal D}}^{\mp}e^{\partial},
\label{comrel}
\end{equation}
which are obviously consistent with 
the commutation relations~(\ref{comrulshift})
by way of $D_{\pm}={\cal D}_{\pm}+ {\overline {\cal D}}^{\pm}$~(\ref{algnn4}).
{}From these relations one can easily see that fermionic covariant
derivatives commute with the shifts by {\it even\/} number of lattice
points only. Therefore, the chirality
constraints (\ref{N=4constr}) are invariant with respect to shifts by
only an {\it even\/} number of lattice points, which is the reason
why only superfields $\{{\cal J}_{2j},{\overline {\cal J}}_{2j}\}$
at {\it even\/} lattice points enter the equations~(\ref{toda11r}). In spite of
this peculiarity the numbers of independent dynamical degrees of freedom
entering the $N=(1|1)$ equations~(\ref{toda11}) and the $N=(2|2)$ 
equations~(\ref{toda11r}) are the same, and they are in one-to-one 
correspondence.

Second, using eqs.~(\ref{recsolution}) and (\ref{basis}), after obvious
manipulations the functionals ${\alpha^{(2p)\pm}_{0,j}} -
{\alpha^{(2p)\pm}_{0,j-1}}$ can identically be represented in the following
form:
\begin{equation}\arraycolsep=0em
\begin{array}{l}
\displaystyle {\alpha^{(2p)\pm}_{0,j}}
- {\alpha^{(2p)\pm}_{0,j-1}}  = \left(1-
e^{-\sum\limits^{2p}_{n=1}{\partial}_n}\right){\alpha^{\pm}_{0,j}}
\vspace{2mm}\\
\displaystyle \qquad  =\pm
\left(1-e^{-2\sum\limits^{2p}_{n=1}{\partial}_n}\right)
\left(1-e^{-2p{\partial}_{2p}}\right) 
 \left\{\prod\limits^{p-1}_{k=1}\left(1-
e^{-\left(2k{\partial}_{2k}+\sum\limits^{2p}_{n=2k+1}{\partial}_n\right)}
\right) \right.
\vspace{2mm}\\
\displaystyle \qquad \left. \times
\left(1+ e^{-\left((2k+1){\partial}_{2k+1}+
\sum\limits^{2p}_{n=2(k+1)}{\partial}_n\right)}\right)\right\}
\left(\prod\limits^{2p}_{m=1} D^{-1}_{\mp}
{\overline {\cal J}}_{j+m}\right).
\end{array}
\label{eq1}
\end{equation}
Then, we find explicitly a product of the expressions inside the first 
two brackets in the second line of eq.~(\ref{eq1}) and use the identity
\begin{equation}
\arraycolsep=0em
\begin{array}{l}
\displaystyle 
{\cal P}_{k-1,k}e^{-\left((k-1){\partial}_{k-1}+
\sum\limits^{2p}_{n=k}{\partial}_n\right)} \left(\prod\limits^{2p}_{m=1}
D^{-1}_{\mp}{\overline {\cal J}}_{j+m}\right) 
\vspace{2mm}\\
\displaystyle \qquad  \qquad \qquad =
e^{-\left(k{\partial}_{k}+\sum\limits^{2p}_{n=k+1}{\partial}_n\right)}
\left(\prod\limits^{2p}_{m=1} D^{-1}_{\mp}{\overline {\cal J}}_{j+m}\right),
\end{array}
\label{identity}
\end{equation}
where ${\cal P}_{k-1,k}$ is a permutation
operator which acts according the rule
\begin{equation}\arraycolsep=0em
\begin{array}{l}
\displaystyle  {\cal P}_{k-1,k}
e^{l{\partial}_k}=
e^{l{\partial}_{k-1}}{\cal P}_{k-1,k}, \qquad
{\cal P}_{k-1,k}e^{l{\partial}_{k-1}}=
e^{l{\partial}_{k}}{\cal P}_{k-1,k}, 
\vspace{2mm}\\
\displaystyle 
{\cal P}_{k-1,k}
\left(\prod\limits^{k-2}_{m=1} D^{-1}_{\mp}{\overline {\cal J}}_{j+m}\right)
D^{-1}_{\mp}{\overline {\cal J}}_{l} D^{-1}_{\mp}
{\overline {\cal J}}_{n} \left(\prod\limits^{2p}_{m=k+1}
D^{-1}_{\mp}{\overline {\cal J}}_{j+m}\right)
\vspace{2mm}\\
\displaystyle  \qquad =\left(\prod\limits^{k-2}_{m=1}
D^{-1}_{\mp}{\overline {\cal J}}_{j+m}\right)
D^{-1}_{\mp}{\overline {\cal J}}_{n}
D^{-1}_{\mp} {\overline {\cal J}}_{l}
\left(\prod\limits^{2p}_{m=k+1} D^{-1}_{\mp}
{\overline {\cal J}}_{j+m}\right).
\end{array}
\label{rule2}
\end{equation}
Equation (\ref{eq1}) now reads
\begin{equation}\arraycolsep=0em
\begin{array}{l}
\displaystyle {\alpha^{(2p)\pm}_{0,j}}
- {\alpha^{(2p)\pm}_{0,j-1}} = \pm
\left(1-e^{-2\sum\limits^{2p}_{n=1}{\partial}_n}\right)
\left(1-e^{-2p{\partial}_{2p}}\right)
\vspace{2mm}\\
\displaystyle \qquad \times
\left\{\prod\limits^{p-1}_{k=1}\left(1-
e^{-2\left(k{\partial}_{2k}+
(k+1){\partial}_{2k+1}+
\sum\limits^{2p}_{n=2(k+1)}{\partial}_n\right)} +
P_{2k,2k+1}\right)\right\}
\left(\prod\limits^{2p}_{m=1} D^{-1}_{\mp}
{\overline {\cal J}}_{j+m}\right) ,
\end{array}\hspace{-10mm}
\label{eq2}
\end{equation}
with
\begin{equation}
P_{2k,2k+1} :=
({\cal P}_{2k,2k+1}-1) e^{-\left(2k{\partial}_{2k}+
\sum\limits^{2p}_{n=2k+1}{\partial}_n\right)}.
\label{eqq2}
\end{equation}
A simple inspection of this formula shows that
for the validity of the chirality constraints~(\ref{constralpha})
it suffices that the functionals
\begin{equation}\arraycolsep=0em
\begin{array}{l}
\label{functionals}
\displaystyle {\cal F}^{(2p)\pm}_j := 
\left(\prod\limits^{2p}_{m=1} D^{-1}_{\mp}
{\overline {\cal J}}_{j+m}\right) ,
\vspace{2mm}\\
\displaystyle {\cal F}^{(l;2p)\pm}_j := 
P_{2l,2l+1} \left(\prod\limits^{2p}_{m=1}
D^{-1}_{\mp} {\overline {\cal J}}_{j+m}\right)=
P_{2l,2l+1}{\cal F}^{(2p)\pm}_j, \quad 1\leq l\leq p -1, 
\vspace{2mm}\\
\displaystyle {\cal F}^{(kl;2p)\pm}_j :=  P_{2k,2k+1} P_{2l,2l+1}
\left(\prod\limits^{2p}_{m=1} D^{-1}_{\mp}
{\overline {\cal J}}_{j+m}\right)=
P_{2k,2k+1}{\cal F}^{(l;2p)\pm}_j, 
\vspace{2mm}\\
\qquad \qquad 
1\leq k < l \leq p -1,\\ 
\cdots\cdots\cdots \cdots\cdots\cdots\cdots \cdots\cdots\cdots \cdots
\vspace{1mm}\\
\displaystyle {\cal F}^{(m\ldots kl;2p)\pm}_j :=
\left(\prod\limits^{p-1}_{k=1} P_{2k,2k+1}\right)
\left(\prod\limits^{2p}_{m=1} D^{-1}_{\mp} {\overline {\cal J}}_{j+m}\right)=
P_{2m,2m+1}{\cal F}^{( \ldots kl;2p)\pm}_j ,
\vspace{2mm}\\
\qquad \qquad 1 \leq m < \cdots < k < l \leq p -1
\end{array}
\end{equation}
appearing in (\ref{eq2})
satisfy the same chirality constraints\footnote{Let us remark that this is
not a necessary condition because the superfields
$\{{\cal J}_{2j},{\overline {\cal J}}_{2j}\}$ at different lattice points
are not linearly independent due to the $N=(2|2)$ Toda lattice equations
(\ref{toda11r}) which relate them.}, i.e.
\begin{equation}
{\overline {\cal D}}^{\mp} {\cal F}^{(m\ldots l;2p)\pm}_{2j}=0, \qquad
{\cal D}_{\mp} {\cal F}^{(m\ldots l;2p)\pm}_{2j+1}=0.
\label{constralpha1}
\end{equation}
Indeed, ${\alpha^{(2p)\pm}_{0,j}} -{\alpha^{(2p)\pm}_{0,j-1}}$ (\ref{eq2})
is a linear functional of ${\cal F}^{m\ldots l(2p)\mp}_j$.
The latter can be shifted with lattice shift operators, but only by an even
number of lattice points, which does not change the chirality
properties of the functionals ${\cal F}^{m\ldots l(2p)\pm}_j$.

In trying to verify that the functionals ${\cal F}^{m\ldots l(2p)\pm}_j$
(\ref{functionals}) do in fact satisfy the conditions~(\ref{constralpha1}),
we substitute $D_{\pm}= {\cal D}_{\pm}+ {\overline {\cal D}}^{\pm}$
(\ref{algnn4}) into eqs.~(\ref{functionals}) 
and use the relations (\ref{N=4constr})
and~(\ref{algnn4}) in order to simplify the resulting expressions.
Let us discuss the outcome:

The functionals ${\cal F}^{(2p)\mp}_{2j}$
and ${\cal F}^{(2p)\mp}_{2j+1}$ become
\begin{equation}\arraycolsep=0em
\begin{array}{l}
\displaystyle 
{\cal F}^{(2p)\mp}_{2j} =\left(\prod\limits^{p}_{m=1}
{\overline {\cal D}}^{\mp}{\partial}^{-1}_{\mp}
{\overline {\cal J}}_{2j+2m-1} {\cal D}_{\mp}{\partial}^{-1}_{\mp}
{\overline {\cal J}}_{2j+2m} \right)=
{\overline {\cal D}}^{\mp}{\partial}^{-1}_{\mp}
{\overline {\cal J}}_{2j+1}{\cal F}^{(2(l-1))\mp}_{2j+1}
{\cal D}_{\mp}
\vspace{3mm}\\
\displaystyle \qquad \times {\partial}^{-1}_{\mp}
{\overline {\cal J}}_{2j+2l}
{\overline {\cal D}}^{\mp}{\partial}^{-1}_{\mp}
{\overline {\cal J}}_{2j+2l+1}
{\cal F}^{(2(p-l-1))\mp}_{2j+2l+1}
{\cal D}_{\mp}{\partial}^{-1}_{\mp}
{\overline {\cal J}}_{2j+2p}, 
\vspace{3mm}\\
\displaystyle {\cal F}^{(2p)\mp}_{2j+1} =
\left(\prod\limits^{p}_{m=1}
{\cal D}_{\mp}{\partial}^{-1}_{\mp}{\overline {\cal J}}_{2j+2m}
{\overline {\cal D}}^{\mp}{\partial}^{-1}_{\mp}
{\overline {\cal J}}_{2j+2m+1} \right)=
{\cal D}_{\mp}{\partial}^{-1}_{\mp}
{\overline {\cal J}}_{2j+2}{\cal F}^{(2(l-1))\mp}_{2j+2}
{\overline {\cal D}}^{\mp}
\vspace{3mm}\\
\displaystyle \qquad \times {\partial}^{-1}_{\mp}
{\overline {\cal J}}_{2j+2l+1}
{\cal D}_{\mp}{\partial}^{-1}_{\mp}
{\overline {\cal J}}_{2j+2l+2}
{\cal F}^{(2(p-l-1))\mp}_{2j+2l+2}
{\overline {\cal D}}^{\mp}{\partial}^{-1}_{\mp}
{\overline {\cal J}}_{2j+2p+1}
\end{array}
\label{fff1}
\end{equation}
and satisfy manifestly the conditions (\ref{constralpha1}),
due to $({\overline {\cal D}}^{\mp})^2=0=({\cal D}_{\mp})^2$.

It turns out that the product of the operators $P_{2k,2k+1}\ldots
P_{2l,2l+1}$  ($1\leq k< \ldots < l \leq p - 1$)
does not change the chirality properties of
${\cal F}^{(2p)\mp}_{2j}$ or ${\cal F}^{(2p)\mp}_{2j+1}$
when applied to these functionals.
Hence, all the functionals
${\cal F}^{m\ldots l(2p)\mp}_j$ (\ref{functionals})
possess the same chirality properties~(\ref{constralpha1}) as
${\cal F}^{(2p)\mp}_{2j}$ resp. ${\cal F}^{(2p)\mp}_{2j+1}$.  In order to
illustrate this fact, let us present the functionals ${\cal
F}^{(l;2p)\mp}_{2j}$ and ${\cal F}^{(l;2p)\mp}_{2j+1}$ (\ref{functionals}),
\begin{equation}\arraycolsep=0em
\begin{array}{l}
\displaystyle {\cal F}^{(l;2p)\mp}_{2j} =
P_{2l,2l+1}{\cal F}^{(2p)\pm}_{2j} =
 {\overline {\cal D}}^{\mp}{\partial}^{-1}_{\mp}
{\overline {\cal J}}_{2j+1}
{\cal F}^{(2(l-1))\mp}_{2j+1}
{\cal D}_{\mp}{\partial}^{-1}_{\mp}
\Big({\overline {\cal J}}_{2j+2l}
{\cal D}_{\mp}{\partial}^{-1}_{\mp}
{\overline {\cal J}}_{2j}
\vspace{3mm}\\
\displaystyle \qquad -
{\overline {\cal J}}_{2j}
{\cal D}_{\mp}{\partial}^{-1}_{\mp}
{\overline {\cal J}}_{2j+2l}\Big)
 {\cal F}^{(2(p-k-1))\mp}_{2j+2l}
{\overline {\cal D}}^{\mp}{\partial}^{-1}_{\mp}
{\overline {\cal J}}_{2j+2p-1}, 
\vspace{3mm}\\
\displaystyle {\cal F}^{(l;2p)\mp}_{2j+1} =
P_{2l,2l+1}{\cal F}^{(2p)\pm}_{2j+1} = {\cal D}_{\mp}{\partial}^{-1}_{\mp}
{\overline {\cal J}}_{2j+2} {\cal F}^{(2(l-1))\mp}_{2j+2}
{\overline {\cal D}}^{\mp}
{\partial}^{-1}_{\mp}
\Big({\overline {\cal J}}_{2j+2l+1}
{\overline {\cal D}}^{\mp}{\partial}^{-1}_{\mp}
{\overline {\cal J}}_{2j+1}
\vspace{3mm}\\
\displaystyle \qquad -
{\overline {\cal J}}_{2j+1}
{\overline {\cal D}}^{\mp}{\partial}^{-1}_{\mp}
{\overline {\cal J}}_{2j+2l+1}\Big) 
{\cal F}^{(2(p-l-1))\mp}_{2j+2l+1}
{\cal D}_{\mp}{\partial}^{-1}_{\mp}
{\overline {\cal J}}_{2j+2p}.
\end{array}\hspace{-10mm}
\label{ffff2}
\end{equation}
Again, they obviously satisfy the conditions (\ref{constralpha1}).

One important remark is in order. When calculating
eqs.~(\ref{ffff2}) we have essentially used the following
important identities,
\begin{equation}\arraycolsep=0em
\begin{array}{l}
\displaystyle {\overline {\cal D}}^{\mp} \left({\overline {\cal J}}_{2j+2l}
{\cal D}_{\mp}{\partial}^{-1}_{\mp} {\overline {\cal J}}_{2j}-
{\overline {\cal J}}_{2j} {\cal D}_{\mp}{\partial}^{-1}_{\mp}
{\overline {\cal J}}_{2j+2l}\right){\overline {\cal D}}^{\mp} = 0,
\vspace{3mm}\\
\displaystyle {\cal D}_{\mp}\left({\overline {\cal J}}_{2j+2l+1}
{\overline {\cal D}}^{\mp}{\partial}^{-1}_{\mp}
{\overline {\cal J}}_{2j+1}-
{\overline {\cal J}}_{2j+1}
{\overline {\cal D}}^{\mp}{\partial}^{-1}_{\mp}
{\overline {\cal J}}_{2j+2l+1}\right) {\cal D}_{\mp}=0,
\end{array}
\label{identities}
\end{equation}
which can easily be checked using the chirality constraints (\ref{N=4constr})
and the algebra of the fermionic covariant derivatives ${\cal D}_{\mp}$ and
${\overline {\cal D}}^{\mp}$~(\ref{algnn4}). Actually, structures like
${\overline {\cal D}}^{\mp} {\overline {\cal J}}_{2j+2l}
{\cal D}_{\mp}$ $\times{\partial}^{-1}_{\mp} {\overline {\cal J}}_{2j}$
and ${\cal D}_{\mp}{\overline {\cal J}}_{2j+2l+1}
{\overline {\cal D}}^{\mp}{\partial}^{-1}_{\mp}
{\overline {\cal J}}_{2j+1}$ appear in eqs.~(\ref{ffff2}) and seem to destroy
the chirality properties we are asking for. However, due to the presence of
the important projector $({\cal P}_{2k,2k+1}-1)$ in the operator
$P_{2k,2k+1}$~(\ref{eqq2}),
these structures enter eqs.~(\ref{ffff2}) only in the particular
combination occuring on the left hand sides of eqs.~(\ref{identities})
and thus disappear.
So, one might say that
${\cal F}^{(l;2p)\mp}_{2j}$ and ${\cal F}^{(l;2p)\mp}_{2j+1}$
owe their chirality properties (\ref{constralpha1}) to the projector
$({\cal P}_{2k,2k+1}-1)$ in eq.~(\ref{eqq2}) 
and the identities~(\ref{identities}).

One can straightforwardly verify that the eqs.~(\ref{ffff2})
 coincide with the equations that can be derived directly
from the equations~(\ref{fff1}) by applying to them the operator
$P_{2l,2l+1}$ and by using the intertwining relations~(\ref{comrel}).
Therefore, eqs.~(\ref{fff1}) and (\ref{ffff2}) are consistent with the
intertwining relations~(\ref{comrel}).
Comparison of eqs.~(\ref{ffff2}) and eqs.~(\ref{fff1}) shows that the operator
$P_{2l,2l+1}$ preserves the chirality of the first $2l$ terms in the
products of
eqs.~(\ref{fff1}) but flips the chirality of the remaining $p{-}2l$ factors.
Moreover, one can easily see from its definition~(\ref{eqq2}) that
$P_{2l,2l+1}$ always preserves the chirality of the first term in a product.
Therefore, the chirality properties of the functionals
$\{{\cal F}^{(l;2p)\mp}_{2j}, {\cal F}^{(l;2p)\mp}_{2j+1}\}$
are identical to those of
$\{{\cal F}^{(2p)\mp}_{2j}, {\cal F}^{(2p)\mp}_{2j+1}\}$.
Furthermore, it is obvious by induction
that the same arguments can be applied to each
functional from the set~(\ref{functionals}) since they are recursively related
by the operator $P_{2l,2l+1}$. Thus, we are led to the conclusion
that the functionals ${\cal F}^{m\ldots l(2p)\mp}_j$~(\ref{functionals})
in fact satisfy the chirality constraints~(\ref{constralpha1}),
which in turn implies that
${\alpha^{(2p)\pm}_{0,j}}-{\alpha^{(2p)\pm}_{0,j-1}}$
satisfy the constraints~(\ref{constralpha}).
This concludes our proof (of half the conjecture).

\section{Further compelling evidence}

Unfortunately, we are unable 
to prove the remaining constraints
(\ref{constralpha0}) for ${\alpha^{(2p)\pm}_{0,j}}-{\alpha^{(2p)\pm}_{0,j-1}}$
and establish our {\it conjecture\/} beyond any doubt.
Short of that, we have explicitly verified the
{\it conjecture\/} for the first three flows
$\frac{\partial}{\partial t^{+}_l}$ 
resulting from the equations (\ref{evolb1r}).
After rather tedious calculations,
the flows can be represented in the following form:
\begin{equation}
\frac{\partial}{\partial t^{+}_1} {\cal J} = {\partial}_+ {\cal J},
\qquad \frac{\partial}{\partial t^{+}_1}{\overline {\cal J}} =
{\partial}_+ {\overline {\cal J}},
\label{eqs2jN=4t1}
\end{equation}
\begin{equation}\arraycolsep=0em
\begin{array}{l}
\displaystyle \frac{\partial}{\partial t^{+}_2} {\cal J} =
-{\partial}^{2}_+ {\cal J} -
{\cal D}_{+}{\cal D}_{-}\left[2{\partial}_+({\cal J} {\partial}^{-1}_-
{\overline {\cal J} })- ({\overline {\cal D}}^{+}
{\overline {\cal D}}^{-}{\partial}^{-1}_-{\cal J} )^2\right],
\vspace{2mm}\\
\displaystyle \frac{\partial}{\partial t^{+}_2}{\overline {\cal J}} =
+{\partial}^{2}_+ {\overline {\cal J}}  - {\overline {\cal D}}^{+}
{\overline {\cal D}}^{-}
\left[2{\partial}_+({\overline {\cal J}}{\partial}^{-1}_-{\cal J})-
({\cal D}_{+}{\cal D}_{-}{\partial}^{-1}_-{\overline {\cal J}})^2\right],
\label{eqs2jN=4}
\end{array}
\end{equation}
\begin{equation}\arraycolsep=0em
\begin{array}{l}
\displaystyle \frac{\partial}{\partial t^{+}_3} {\cal J} =
{\partial}^{3}_+ {\cal J} +{\cal D}_{+}{\cal D}_{-}\Bigl\{3
{\partial}_+ \Bigl[
({\partial}_+{\cal J}){\partial}^{-1}_-{\overline {\cal J}} +
({\cal J}{\partial}^{-1}_-{\overline {\cal J}})
{\cal D}_{+} {\cal D}_{-}{\partial}^{-1}_-{\overline {\cal J}}
\vspace{2mm}\\
\displaystyle 
\qquad \qquad
-\frac{1}{2}({\overline {\cal D}}^{+}{\overline {\cal D}}^{-}
{\partial}^{-1}_-{\cal J})^2\Bigr]
-({\overline {\cal D}}^{+}{\overline {\cal D}}^{-}{\partial}^{-1}_-
{\cal J})^3 -3({\overline {\cal D}}^{+}{\overline {\cal D}}^{-}
{\partial}^{-1}_-{\cal J})^2 {\cal D}_{+}{\cal D}_{-}
{\partial}^{-1}_-{\overline {\cal J}} 
\vspace{2mm}\\
\displaystyle 
\qquad \qquad
 + 3{\cal J}({\partial}^{-1}_- {\partial}_+ {\overline {\cal J}})
~{\overline {\cal D}}^{+} {\overline {\cal D}}^{-}{\partial}^{-1}_-
{\cal J} +3{\cal J}{\partial}^{-1}_- {\partial}_+
({\overline {\cal J}} ~{\overline {\cal D}}^{+}
{\overline {\cal D}}^{-}{\partial}^{-1}_- {\cal J}) 
\vspace{2mm}\\
\displaystyle
\qquad \qquad
+3({\overline {\cal D}}^{+} {\cal J})({\overline {\cal D}}^{-}
{\partial}^{-1}_- {\partial}_+
{\cal J}){\partial}^{-1}_-{\overline {\cal J}}
-3({\overline {\cal D}}^{+} {\cal J}){\overline {\cal D}}^{-}
{\partial}^{-1}_- {\partial}_+ ({\cal J}{\partial}^{-1}_-
{\overline {\cal J}}) \Bigr\},
\vspace{2mm}\\
\displaystyle \frac{\partial}{\partial t^{+}_3}{\overline {\cal J}} =
{\partial}^{3}_+ {\overline {\cal J}}  +
{\overline {\cal D}}^{+}{\overline {\cal D}}^{-}\Bigl\{3{\partial}_+
\Bigl[-({\partial}_+{\overline {\cal J}}){\partial}^{-1}_-{\cal J} +
({\overline {\cal J}}{\partial}^{-1}_-{\cal J})
{\overline {\cal D}}^{+} {\overline {\cal D}}^{-}{\partial}^{-1}_-{\cal J}
\vspace{2mm}\\
\displaystyle 
\qquad \qquad
 +\frac{1}{2}({\cal D}_{+}{\cal D}_{-}
{\partial}^{-1}_-{\overline {\cal J}})^2\Bigr]
-({\cal D}_{+}{\cal D}_{-}{\partial}^{-1}
{\overline {\cal J}})^3 -3
({\cal D}_{+}{\cal D}_{-}{\partial}^{-1}_-{\overline {\cal J}})^2
{\overline {\cal D}}^{+}{\overline {\cal D}}^{-} {\partial}^{-1}_-
{\cal J} 
\vspace{2mm}\\
\displaystyle 
\qquad \qquad
 +3{\overline {\cal J}}({\partial}^{-1}_- {\partial}_+ {\cal J})
~{\cal D}_{+} {\cal D}_{-}{\partial}^{-1}_-
{\overline {\cal J}} +3{\overline {\cal J}}
{\partial}^{-1}_- {\partial}_+ ({\cal J} {\cal D}_{+}
{\cal D}_{-}{\partial}^{-1}_- {\overline {\cal J}}) 
\vspace{2mm}\\
\displaystyle 
\qquad \qquad
 +3({\cal D}_{+} {\overline {\cal J}})({\cal D}_{-}
{\partial}^{-1}_- {\partial}_+
{\overline {\cal J}}){\partial}^{-1}_-{\cal J}
-3({\cal D}_{+} {\overline {\cal J}}){\cal D}_{-}
{\partial}^{-1}_- {\partial}_+ ({\overline {\cal J}}
{\partial}^{-1}_-{\cal J}) \Bigr\},
\end{array}
\label{eqs2jN=4t3}
\end{equation}
where ${\cal J}\equiv
{\cal J}_{2j}(x^+,\theta^+,\eta^+;x^-,\theta^-,\eta^-)$ and
${\overline {\cal J}} \equiv {\overline {\cal J}}_{2j}
(x^+,\theta^+,\eta^+;x^-,\theta^-,\eta^-)$.
The right hand sides are obviously consistent
with the chirality constraints~(\ref{N=4constr}).

Considering the proved first half 
(i.e. eqs.~(\ref{constralpha})) of the conjectured chirality constraints 
(\ref{constralpha})--(\ref{constralpha0})
and the proof of the {\it conjecture\/} for the one-dimensional reduction
to the $N=4$ supersymmetric Toda chain hierarchy~\cite{ds1},
it is reasonable to believe that our {\it conjecture\/} is valid for the
whole hierarchy of flows $\frac{\partial}{\partial t^{\pm}_l}$.

\section{Five real forms}
Direct verification shows that the flows (\ref{eqs2jN=4t1})--(\ref{eqs2jN=4t3})
admit the following five
inequivalent\footnote{
We mean that it is not possible to relate them via obvious symmetries,
perhaps, some elusive equivalence exists, nevertheless, cf.~\cite{serganova}.}
complex conjugations:
\begin{equation}
({\cal J},{\overline {\cal J}})^{*}=-({\cal J},{\overline {\cal J}}),
\qquad (x^{\pm},{{\theta}^{\pm}},\eta^{\pm})^{*}=
(-x^{\pm},{\theta}^{\pm},-\eta^{\pm}),
\qquad (t^{\pm}_l)^{*}=(-1)^{l}t^{\pm}_l,\hspace{-4mm}
\label{conj1}
\end{equation}
\begin{equation}\arraycolsep=0em
\begin{array}{l}
({\cal J},{\overline {\cal J}})^{\bullet}=
(~{\cal J}- {\cal D}_-{\cal D}_+\ln
{\overline {\cal J}},~{\overline {\cal J}}),
\vspace{2mm}\\
 (x^{\pm},{{\theta}^{\pm}},\eta^{\pm})^{\bullet}=
(-x^{\pm},{\theta}^{\pm},-\eta^{\pm}),
\qquad
(t^{\pm}_l)^{\bullet}=-t^{\pm}_l,
\end{array}
\label{conj2}
\end{equation}
\begin{equation}
({\cal J},{\overline {\cal J}})^{\star}
=({\overline {\cal J}},~{\cal J}),
\qquad (x^{\pm},{{\theta}^{\pm}},\eta^{\pm})^{\star}=
(-x^{\pm},{\theta}^{\pm},\eta^{\pm}),
\qquad (t^{\pm}_l)^{\star}=-t^{\pm}_l,
\label{conj3}
\end{equation}
\begin{equation}
({\cal J},{\overline {\cal J}})^{\dagger}=-({\cal J},
{\overline {\cal J}}), \qquad
(x^{\pm},{{\theta}^{\pm}},\eta^{\pm})^{\dagger}=(-x^{\pm},
i{\eta}^{\pm},i\theta^{\pm}),
\qquad (t^{\pm}_l)^{\dagger}=(-1)^{l}t^{\pm}_l,\hspace{-4mm}
\label{conj1new}
\end{equation}
\begin{equation}
({\cal J},{\overline {\cal J}})^{\ddagger}=({\cal J},
{\overline {\cal J}}), \qquad
(x^{\pm},{{\theta}^{\pm}},\eta^{\pm})^{\ddagger}=
(-x^{\mp},{\theta}^{\mp},-\eta^{\mp}),
\qquad (t^{\pm}_l)^{\ddagger}=(-1)^{l}t^{\mp}_l.
\label{conj2new}
\end{equation}
These involutions extract five inequivalent real forms of the hierarchy.
In particular, the flows of the real form corresponding to the conjugation
(\ref{conj3}) reproduce the algebra of {\it real\/} $N=(2|2)$ supersymmetry.
We use the standard convention regarding complex
conjugation of products involving odd operators and functions
(see, e.g., the books~\cite{ggrs}). In particular, if ${\mathbb D}$ is some
even differential operator acting on a superfield $F$, we define the
complex conjugate of ${\mathbb D}$ by
$({\mathbb D}F)^*={{\mathbb D}}^*F^*$.

A combination of the two complex conjugations (\ref{conj3}) and (\ref{conj2}), 
when applied twice, generates a manifestly $N=(2|2)$ supersymmetric form of the
$N=(2|2)$ Toda lattice equations,~(\ref{toda11r}):
\begin{equation}
{\cal J}^{\star \bullet \star \bullet}=
{\cal J} -{\cal D}_-{\cal D}_+\ln {\overline {\cal J}}, \qquad
{\overline {\cal J}}^{\star \bullet \star \bullet}={\overline {\cal J}} -
{\overline {\cal D}}^-{\overline {\cal D}}^+
\ln {\cal J}^{\star \bullet \star\bullet}.
\label{conj2jn=4}
\end{equation}
In other words, if the set $\{{\cal J},{\overline {\cal J}} \}$
is a solution of the $N=(2|2)$ Toda lattice hierarchy, then the set
$\{{\cal J}^{\star \bullet \star \bullet},
{\overline {\cal J}}^{\star \bullet \star \bullet}\}$, related to
the former by eqs.~(\ref{conj2jn=4}), is also a solution of the hierarchy.

Finally, a combination of the two complex conjugations
(\ref{conj1}) and (\ref{conj2}) generates a~se\-cond-order discrete
symmetry of the flows $\frac{\partial}{\partial t^{\pm}_{2l+1}}$,
\begin{equation}
({\cal J},{\overline {\cal J}})^{\bullet *}=
-({\cal J} + {\cal D}_-{\cal D}_+\ln
{\overline {\cal J}},{\overline {\cal J}}), \qquad
({\cal J},{\overline {\cal J}})^{\bullet *\bullet *}=
({\cal J},{\overline {\cal J}}).
\label{symm}
\end{equation}

\section{Conclusion}
In this letter we have proposed 
an $N=(2|2)$ superfield formulation
of the $N=(2|2)$ supersymmetric Toda lattice hierarchy and
have constructed five different real forms in $N=(2|2)$ superspace.

\subsection*{Acknowledgments}

A.S. would like to thank the Institut f\"ur Theoretische Physik,
Universit\"at
Hannover for the hospitality during the course of this work.
This work was partially supported by the DFG Grant No.~436 RUS 113/359/0
(R), RFBR-DFG Grant No.~99-02-04022, PICS Project No.~593,
RFBR-CNRS Grant No.~98-02-22034, Nato Grant No. PST.CLG 974874,
RFBR Grant No.~99-02-18417.
O.L. is grateful to the Erwin Schr\"odinger International Institute for
Mathematical Physics in Vienna, where this work was completed.

\newpage


\begin{thebibliography}{20}

\small
\topsep0mm
\partopsep0mm
\parsep0mm
\itemsep0mm

\bibitem{ls}
Leznov A~N and Sorin A~S,
Two-Dimensional Superintegrable Mappings and Integrable
Hierarchies in the $(2|2)$ Superspace,
{\it Phys. Lett.} {\bf B389} (1996), 494, hep-th/9608166;
Integrable Mappings and Hierarchies in the $(2|2)$ Superspace,
{\it Nucl. Phys. B (Proc. Suppl.)} {\bf 56} (1997), 258.
\bibitem{darboux}
Darboux G, Le\c{c}ons sur la th\'eorie g\'en\'erale des surfaces et
des applications g\'eom\'etriques du calcul infinit\'esimal, Paris,
1887--1896.                            
\bibitem{ols}
Lechtenfeld O and Sorin A,
Fermionic Flows and Tau Function of the $N=(1|1)$ Superconformal Toda
Lattice Hierarchy, {\it Nucl. Phys.} {\bf B557} (1999), 535, solv-int/9810009.
\bibitem{dly}
Derjagin V~B, Leznov A~N and Yuzbashyan E~A,
Two-Dimensional Integrable Mappings and Explicit Form of Equations
     of $(1+2)$-Dimensional Hierarchies of Integrable Systems, 
IHEP-95-26, MPI 96-39, 1996.
\bibitem{bx}
Bonora L and Xiong C~S,
 An Alternative Approach to KP Hierarchy in Matrix Models,
{\it Phys. Lett.} {\bf B285} (1992), 191, hep-th/9204019;
 Matrix Models without Scaling Limit,
{\it Int. J. Mod. Phys.} {\bf A8} (1993), 2973, hep-th/9209041.
\bibitem{lsy}
Leznov A~N, Shabat A~B  and  Yamilov R~I,
Canonical Transformations Generated by Shifts in Nonlinear Lattices,
{\it Phys. Lett.} {\bf A174} (1993), 397.
\bibitem{ly}
Leznov A~N  and  Yuzbashyan E~A,
Integrable Mappings for Non-Commutative Objects, {\it
Rept. Math. Phys.}
{\bf 43} (1999), 207, hep-th/9609024.
\bibitem{eh}
Evans J and Hollowood T, Supersymmetric Toda Field Theories,
{\it Nucl. Phys.} {\bf B352} (1991),  723.
\bibitem{lds}
Derjagin V~B, Leznov A~N and Sorin A, The Solution of the
$N=(0|2)$ Superconformal f-Toda Lattice, {\it Nucl. Phys.} {\bf B527} (1998),
643, solv-int/9803010.
\bibitem{ds1}
Delduc F and Sorin A,
Lax Pair Formulation of the $N=4$ Supersymmetric Toda
Chain (KdV) Hierarchy in $N=4$ Superspace, in preparation.
\bibitem{serganova} Serganova V~V, 
Automorphisms of Lie Superalgebras of String Theories,
{\it Funktional. Anal. i~Prilozhen.} {\bf 19} (1985), 75 (in Russian). 
\bibitem{ggrs}
Gates S~J (Jr.), Grisaru M~T, Ro\c{}ek M and Siegel W,
 Superspace or One Thousand and One Lessons in Supersymmetry,
the Benjamin/Cummings Publishing Company Inc., 1983, 58--59;\\
West P, Introduction to Supersymmetry and Supergravity,
Extended Second Edition,
World Scientific, 1990, 393--394.
\label{Lech-lastpage}

\end{thebibliography}
\end{document}